\documentclass[prd,preprint,nofootinbib,amssymb, amsmath,mathrsfs,nobibnotes,aps,11pt]{revtex4}
\usepackage{color}
\usepackage{graphicx}
\usepackage[colorlinks=true,citecolor=blue,linkcolor=blue]{hyperref}

\newcommand{\nfw}{\textrm{NFW}}
\newcommand{\esc}{\textrm{esc}}
\newcommand{\rel}{\textrm{rel}}
\newcommand{\Vmax}{V_\textrm{max}}
\newcommand{\rmax}{r_\textrm{max}}
\newcommand{\DM}{\textrm{DM}}

\graphicspath{ {../Figures/} }

\begin{document}

\title{Sommerfeld-Enhanced $J$-Factors For Dwarf Spheroidal Galaxies}
\preprint{\hbox{UH-511-1274-2017}}
\preprint{\hbox{CETUP2016-010}}

\author{Kimberly~K.~Boddy$^{1,}$\footnote{kboddy@hawaii.edu}}
\author{Jason Kumar$^{1,}$\footnote{jkumar@hawaii.edu}}
\author{Louis E.~Strigari$^{2,}$\footnote{strigari@tamu.edu}}
\author{Mei-Yu Wang$^{2,}$\footnote{meiyu@physics.tamu.edu}}
\affiliation{$^1$Department of Physics \& Astronomy, University of Hawaii, Honolulu, HI 96822, USA}
\affiliation{$^2$Department of Physics \& Astronomy, Mitchell Institute for Fundamental Physics and Astronomy, Texas A\&M University, College Station, TX 77843, USA}

\begin{abstract}
For models in which dark matter annihilation is Sommerfeld-enhanced, the annihilation cross section increases at low relative velocities.
Dwarf spheroidal galaxies (dSphs) have low characteristic dark matter particle velocities and are thus ideal candidates to study such models.
In this paper we model the dark matter phase space of dSphs as isotropic and spherically-symmetric, and determine the $J$-factors for several of the most important targets for indirect dark matter searches.
For Navarro-Frenk-White density profiles, we quantify the scatter in the $J$-factor arising from the astrophysical uncertainty in the dark matter potential.
We show that, in Sommerfeld-enhanced models, the ordering of the most promising dSphs may be different relative to the standard case of velocity-independent cross sections.
This result can have important implications for derived upper limits on the annihilation cross section, or on possible signals, from dSphs.
\end{abstract}
\maketitle

\section{Introduction}

A major strategy for the indirect detection of dark matter is the search for photons arising from dark matter annihilation in dwarf spheroidal galaxies (dSphs)~\cite{Conrad:2015bsa}.
The dark matter halo masses of dSphs are well constrained from stellar kinematics~\cite{McConnachie:2012vd,Walker:2012td,Battaglia:2013wqa,Strigari:2013iaa}, and the systematic uncertainties associated with the expected backgrounds~\cite{Winter:2016wmy} are relatively small compared with other astrophysical targets and channels used in indirect detection.
Null detection results from Fermi-LAT place strong limits on the dark matter annihilation cross section for particles with mass $\lesssim 100$ GeV~\cite{Geringer-Sameth:2014qqa,Ackermann:2015zua}, ruling out thermal relic dark matter for velocity-independent cross sections to some final states.
Targeted ground-based observatories provide the most stringent limits for masses $\gtrsim 1$ TeV~\cite{Ahnen:2016qkx}.

In determining the flux of photons arising from dark matter annihilation, the main astrophysical dependence is encapsulated in the $J$-factor of the target.
If the annihilation cross section $\sigma_A v$ is velocity-independent, the $J$-factor is simply an integral over the line-of-sight and over a given angular region of the square of the dark matter density profile of the target.
With this assumption the $J$-factor is independent of the underlying particle physics such as the dark matter mass and cross section, and furthermore, it is independent of the particular phase space distribution of the dark matter.
Many authors have determined the $J$-factors from the stellar kinematics of dSphs under the assumption of a velocity-independent annihilation cross section~\cite{Essig:2010em,Martinez:2013els,Geringer-Sameth:2014yza,Bonnivard:2015xpq,Evans:2016xwx,Sanders:2016eie}.

However, from a theoretical perspective, the annihilation cross section may be velocity dependent; for example, theoretically well-studied models have $p$-wave suppressed dark matter annihilation ($\sigma_A v \propto v^2$).
Additionally, it has been long appreciated that there are some dark matter models in which dark matter annihilation exhibits a {\it Sommerfeld enhancement} at low relative velocities~\cite{ArkaniHamed:2008qn}.
If the annihilation cross section is velocity-dependent, the photon flux arising from dark matter annihilation does in fact depend on the dark matter velocity distribution, and the astrophysical dependence cannot be entirely factorized from the particle physics~\cite{Robertson:2009bh,Ferrer:2013cla}.
However, there has not yet been a systematic study of $J$-factors for velocity-dependent cross sections.

In this paper we determine the analog of the $J$-factor which is relevant for Sommerfeld-enhanced dark matter annihilation.
We use a simple isotropic and spherically-symmetric model%
\footnote{We do not consider the enhancement effects of sub-substructure~\cite{Pieri:2009zi}, since the results are subject to theoretical extrapolations of the concentration-mass relation.}
for the dark matter phase space distribution of the dSphs, which --- under the assumption of a model for the gravitational potential --- is constrained by the measured stellar velocity distributions.
We show that this new astrophysics factor, denoted as $J_S$, depends on two parameters which are determined by the detailed particle physics of dark matter annihilation.

The results that we present have important practical applications for interpreting limits on (or establishing possible detections of) a dark matter annihilation signal from dSphs.
For example, if we consider a single dSph target, the $J$-factor is the quantity which allows one to translate a statistical bound on (or an observed excess due to) the number of photons arising from dark matter annihilation into a corresponding limit on (or value for) the dark matter annihilation cross section.
Similarly, a determination of the Sommerfeld-enhanced $J$-factor would allow one to translate observations of the photon flux into preferred and/or excluded values of the Sommerfeld-enhanced cross section.

In addition, we may consider the results from a combination of multiple dSph targets.
If one observes multiple targets, the relative ordering of the $J$-factors also provides an important consistency check for the dark matter interpretation of any potentially observed excess.
For a given dark matter annihilation cross section, the expected flux of photons arising from dark matter annihilation scales with the $J$-factor~\cite{Fermi-LAT:2016uux}.
Thus, if an excess is observed in one dwarf spheroidal target, one would also expect excesses in other targets with larger $J$-factors; a failure to see such excesses would draw into question the consistency of the dark matter interpretation of the photon signal.
However, since the different dwarf spheroidal galaxies can have very different velocity dispersions, the relative ordering of the Sommerfeld-enhanced $J$-factors may be quite different from that of the ordinary $J$-factor.
Thus, a pattern of excesses in the gamma-ray emissions of many dwarf spheroidal galaxies, which may appear inconsistent with velocity-independent dark matter annihilation, may still be consistent with Sommerfeld-enhanced dark matter annihilation.

The plan of this paper is as follows.
In section~\ref{sec:distribution}, we review the formalism for obtaining the dark matter velocity distribution from stellar observations.
In section~\ref{sec:sommerfeld}, we review the theoretical considerations underlying the Sommerfeld enhancement of dark matter annihilation and derive an expression for the Sommerfeld-enhanced $J$-factor, $J_S$.
In section~\ref{sec:results} we present our results for $J_S$ for several dwarf spheroidal galaxies.
We conclude with a discussion of our results in section~\ref{sec:conclusion}.

\section{Dark matter distribution function and density profiles}
\label{sec:distribution}

In a dSph, the positions in the plane of the sky and the line-of-sight velocities of stars are resolved, leading to a measurement of the projected stellar velocity distribution.
For the analysis in this paper, we are interested in the 3D dark matter velocity distribution, which is not necessarily the same as the stellar velocity distribution.
To determine this distribution, we use constraints on the gravitational potentials of dSphs, in combination with well-motivated theoretical assumptions.

To calculate the dark matter velocity distribution, we assume the orbits of the dark matter particles are isotropic and the potential is spherically-symmetric.
This approximation is justified when examining satellite galaxies in cosmological simulations in which the star particles have ratios of tangential-to-radial velocity anisotropy in the range $\sim 0.8-1.3$~\cite{Campbell:2016vkb}.
Additional studies of the velocity anisotropy profiles of subhalos in dark matter-only simulations are consistent with this range out to the subhalo virial radius~\cite{Vera-Ciro:2014ita}.
Under these assumptions, we can use the Eddington formula for the isotropic distribution function
\begin{equation}
  f_\DM(\epsilon) = \frac{1}{\sqrt{8}\pi^2} \int_\epsilon^0
  \frac{d^2 \rho_\DM}{d\Psi^2}\frac{d\Psi}{\sqrt{\epsilon - \Psi }} \ ,
  \label{eq:eddington}
\end{equation}
which is a function of energy alone.
Here, $\rho_\DM(r)$ is the dark matter density profile, and the function $\Psi(r) < 0$ is the spherically-symmetric gravitational potential, which depends on the parameters of the dark matter density profile.
The gravitational binding energy per mass of a dark matter particle is $\epsilon = v^2/2 + \Psi(r) < 0$, and $v$ is the modulus of the velocity of a dark matter particle.
Thus, the quantity $f_\DM (\epsilon)$ is implicitly a function of $v$ and $r$ and is equivalent to the velocity distribution function $f(r,v) \equiv f_\DM (\epsilon(r,v))$.
The velocity distribution obeys the normalization
\begin{equation}
  \rho_\DM (r) = 4\pi \int_0^{v_\esc} dv \, v^2 f(r,v) \ ,
\end{equation}
where $v_\esc(r) = \sqrt{-2\Psi(r)}$ is the maximum velocity obtainable for a gravitationally-bound particle at radius $r$.

We assume a Navarro-Frenk-White (NFW) form of the dark matter density profile: $\rho_\nfw (r) = \rho_s/[(r/r_s) (1+r/r_s)^2]$.
The NFW profile can be a cast as a function of the scale density and scale radius $(\rho_s,r_s)$, or the maximum circular velocity and the radius of maximum circular velocity ($\Vmax, \rmax$).
These quantities are related via
\begin{equation}
  \rmax = 2.16~r_s \ , \qquad  \Vmax= 0.465 ~\sqrt{4 \pi G \rho_s r_s^2} \ .
  \label{eq:rmaxvmax}
\end{equation}
Hence, the dark matter velocity distribution $f(r,v)$ depends only on the parameters ($\rho_s,r_s$) of the NFW profile, or ($\Vmax, \rmax$).
The properties of dark matter distributions with cuspy NFW profiles have been studied in e.g.~Refs.~\cite{Widrow:2000,Evans:2005tn}.

The parameters ($\rho_s,r_s$) or ($\Vmax, \rmax$) can be bound by observations of the average stellar line-of-sight velocity distribution for each dSph.
In order to do so, we define the stellar distribution function as $f_\star$.
Again under the assumption of spherical symmetry and isotropy for the stellar distribution, $f_\star$ can be calculated from Eq.~\eqref{eq:eddington} given a stellar density profile, $\rho_\star$, which we take to be a Plummer profile with the best-fit half-light radius $r_h$ for each dSph~\cite{McConnachie:2012vd}.
With our definition of $f_\star$, the average stellar velocity dispersion at a radius $r$ is
\begin{equation}
  \langle \sigma_\star^2 (r)\rangle =
  \frac{\int v_\star^4 f_\star(v_\star,r)\, dv_\star}
       {\int v_\star^2 f_\star(v_\star,r)\, dv_\star} \ ,
  \label{eq:avgsigma}
\end{equation}
where $v_\star$ refers to the velocity of the stars.
To obtain the quantity that most closely approximates the observed projected velocity dispersion averaged over the entire galaxy, we then calculate
\begin{equation}
  3\times\langle \sigma_{\star}^2 \rangle =
  \frac{\int \langle \sigma_{\star}^2(r) \rangle \rho_{\star} dV}
       {\int \rho_{\star} dV} \ .
  \label{eq:avgsigmaprojected}
\end{equation}

\begin{table*}
  {\renewcommand{\arraystretch}{1.15}
    \renewcommand{\tabcolsep}{0.15cm}
    \begin{tabular}{l l c c c c c}
      \hline
      \hline
      dSph & Ref. & $\langle\sigma_\star\rangle$ & $r_h$ & $D$ & $V_{\rm max}$ & $r_{\rm max}$ \\
      & & [km/s] & [kpc] & [kpc] & [km/s] & [kpc] \\
      \hline
      Coma Berenices & \cite{Simon:2007dq}
      & $4.6$ & $0.077$ & $44.0$ & 9.8  & 0.38 \\
      Ursa Minor     & \cite{McConnachie:2012vd}
      & $9.5$ & $0.181$ & $76.0$ & 24.1 & 1.32 \\
      Draco          & \cite{McConnachie:2012vd}
      & $9.1$ & $0.221$ & $76.0$ & 17.7 & 0.86 \\
      Segue 1        & \cite{Simon:2010ek}
      & $3.9$ & $0.029$ & $23.0$ & 16.2 & 0.76 \\
      Reticulum II   & \cite{Simon:2015,Walker:2015mla}
      & $3.3$ & $0.055$ & $32.0$ & 7.6  & 0.28 \\
      \hline
    \end{tabular}
    \medskip
  }
\caption{Properties of five dSphs and NFW profile parameter values.
  Column 2 gives the references for the quantities in columns 3--5: the average stellar velocity dispersion ($ \langle \sigma_{\star} \rangle$), the half-light radius ($r_h$), and the distance to the dSph ($D$).
  The values of ($\Vmax,\rmax$) are derived from the central points for each dSph in Fig.~\ref{fig:rmaxvmaxgrid}.}
\label{tb:orbit}
\end{table*}

\begin{figure}[t]
  \centering
  \includegraphics[scale=0.8]{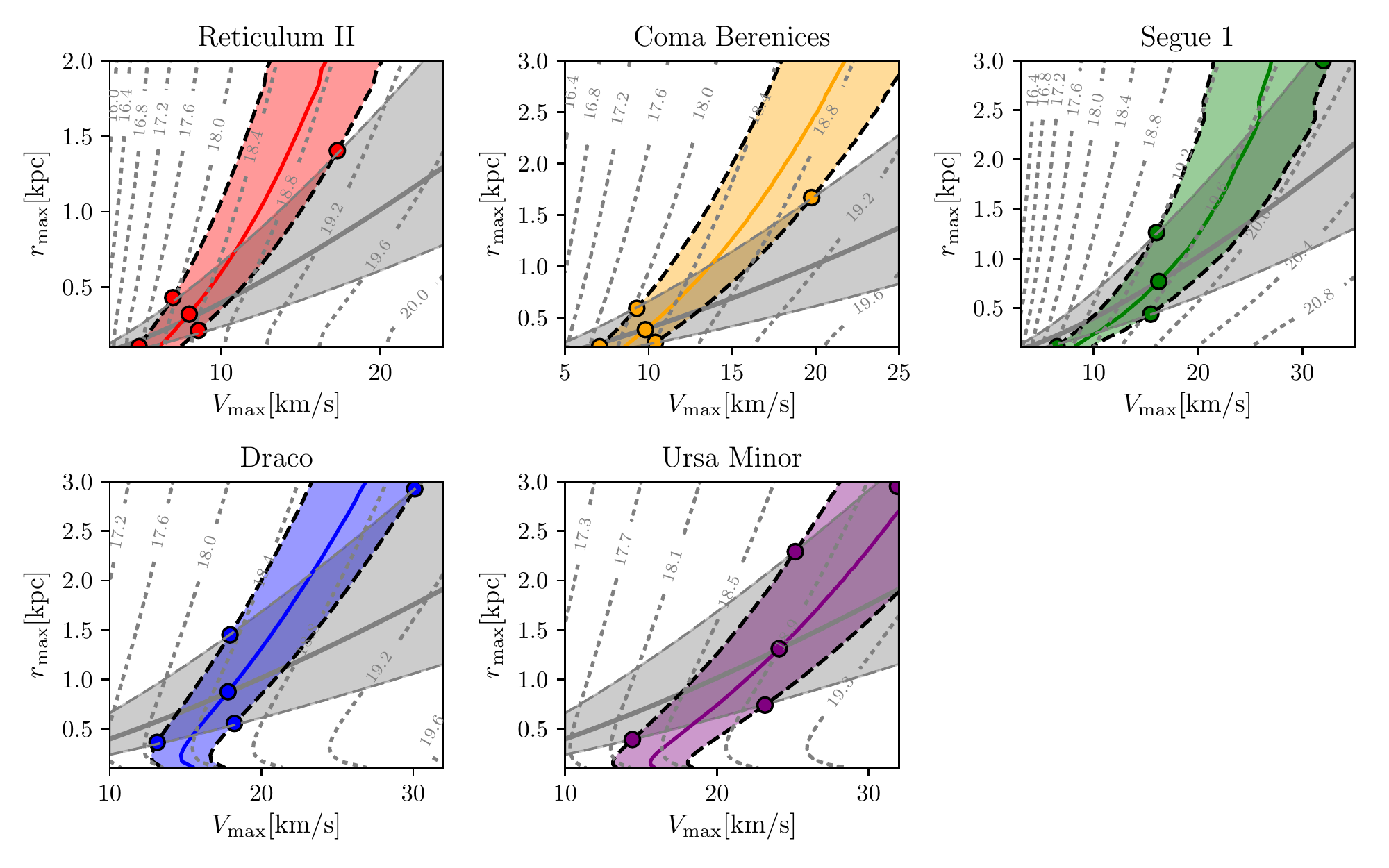}
  \caption{The observable and theoretically-preferred regions in the $\Vmax - \rmax$ parameter space for the five dSphs under consideration.
    Along the solid colored curves are points that match the average stellar velocity dispersion, and the corresponding black dashed curves match the 1$\sigma$ uncertainties.
    The solid gray curves indicate the median ($\Vmax,\rmax$) relation for subhalos~\cite{Martinez:2009jh} in the dark matter-only Aquarius simulation~\cite{Springel:2008cc}, and the dashed gray curves represent the scatter in this relation.
    Dotted gray lines indicate the ordinary $J$-factors, for which the annihilation cross section is assumed to be independent of velocity.
    The filled circles define the points in parameter space that we use in calculating the Sommerfeld-enhanced $J$-factors.}
  \label{fig:rmaxvmaxgrid}
\end{figure}

We use the observed values of $\sigma_{\star}^2$ to constrain the parameters ($\Vmax, \rmax$) of several dSphs.
We focus on five of the most promising dSphs for indirect detection: Segue 1, Reticulum II, Coma Berenices, Draco, and Ursa Minor.
The stellar velocity dispersions, half-light radii, and the distance to the dSph are summarized in Table~\ref{tb:orbit}.
For these five dSphs, which have the largest ordinary $J$-factors, Fig.~\ref{fig:rmaxvmaxgrid} shows the ($\Vmax, \rmax$) parameter space that is consistent with the observed average velocity dispersion and its measured uncertainty.

\begin{figure}[t]
  \centering
  \includegraphics[scale=0.8]{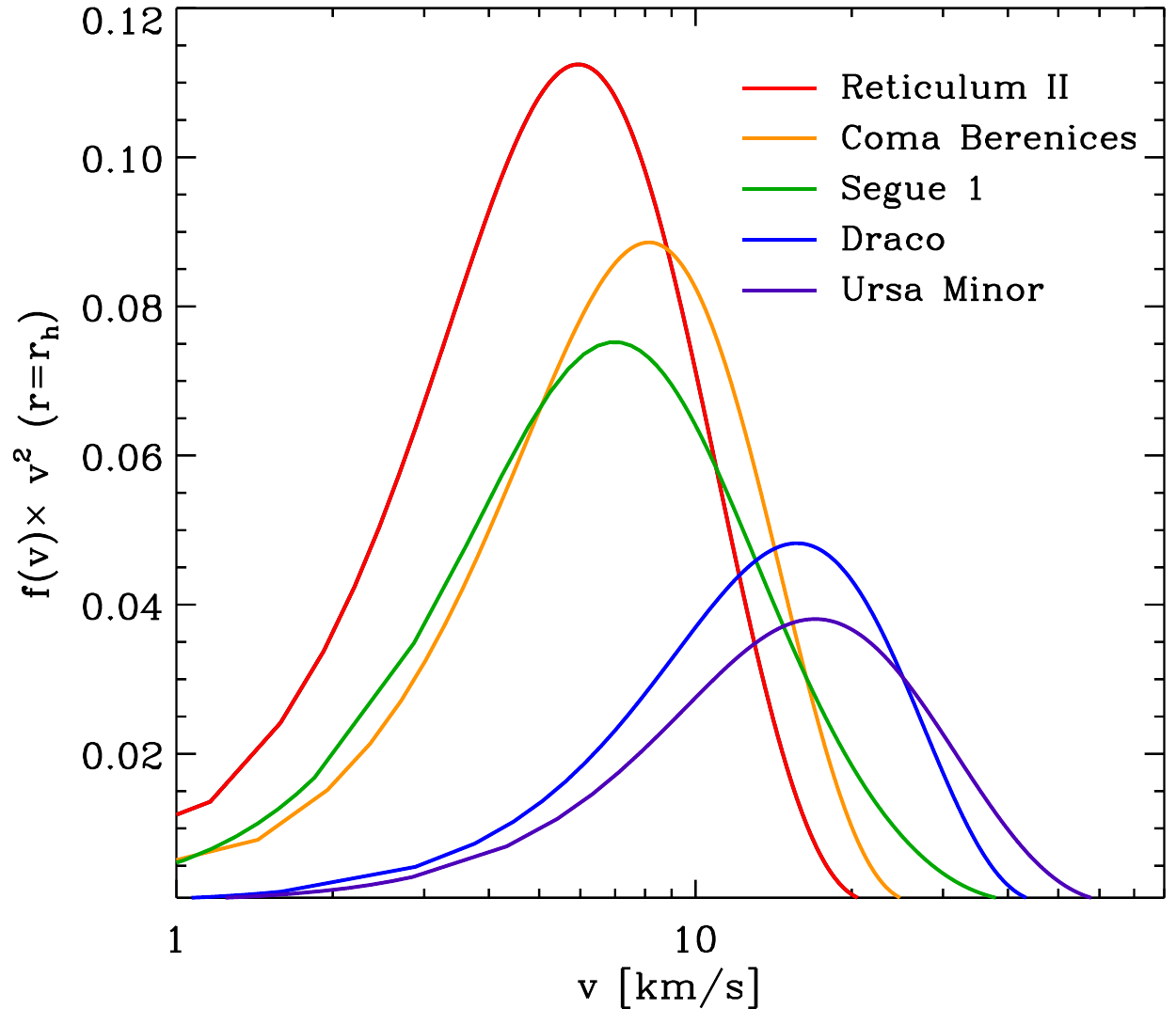}
  \caption{Plots of $v^2 f(v)$, evaluated at the half-light radius, for the five dSphs using the central point of Fig.~\ref{fig:rmaxvmaxgrid}.
    All curves are normalized as $\int v^2 f(v) dv = 1$.}
  \label{fig:fv_dSph}
\end{figure}

The range of ($\Vmax,\rmax$) parameter space can be further bound by appealing to the results from cosmological simulations.
Figure~\ref{fig:rmaxvmaxgrid} also shows the regions consistent with the ($\Vmax,\rmax$) relation for subhalos in the dark matter-only Aquarius simulation (see Fig.~26 of Ref.~\cite{Springel:2008cc}).
Specifically, we adopt the relation $\log(\rmax/\mathrm{kpc}) = 1.35\log[\Vmax/(\mathrm{km/s})]-1.75$ from Eq.~(16) in Ref.~\cite{Martinez:2009jh}, which provides a good description of the Aquarius results, with a uniform scatter of $\sigma_{\log(\rmax/\mathrm{kpc})}=0.22$ for the entire range of $\Vmax$.
To approximate the observational uncertainty in the $J$-factor within the context of our isotropic and spherically-symmetric NFW model, we combine the results from both the measured velocity dispersion and theoretical ($\Vmax , \rmax$) relation.
Specifically, we consider the area in Fig.~\ref{fig:rmaxvmaxgrid} that is defined by the intersection of the observed velocity dispersion band and the subhalo ($\Vmax,\rmax$) band.
From this area, we define five points: the four points that represent the intersection of the outer boundaries of each band, and the central point where the central values of the velocity dispersion and the ($\Vmax,\rmax$) lines cross.
For most of the dSphs, this area has $\Vmax \lesssim 35$ km/s, corresponding to plausible subhalo hosts of dSphs.
The only exception is Ursa Minor, for which we place an upper bound on $\Vmax$ such that $\rmax < 3$ kpc, which corresponds to an estimate for the dark matter tidal radius.
For all dSphs, the central ($\Vmax , \rmax$)  values and stellar velocity dispersions are also listed in Table~\ref{tb:orbit}.
In Fig.~\ref{fig:fv_dSph} we show the velocity distribution at the half-light radius for the five dSphs, using the central point in Fig.~\ref{fig:rmaxvmaxgrid}.

Note that the values of the $J$-factor indicated in Fig.~\ref{fig:rmaxvmaxgrid} assume an NFW profile and isotropic orbits.
These $J$-factors may differ from previous calculations in the literature~\cite{Martinez:2013els,Geringer-Sameth:2014yza,Bonnivard:2015xpq,Evans:2016xwx,Sanders:2016eie}, which allow for non-NFW profiles, anisotropic stellar velocity dispersions, and assume a Gaussian likelihood for the stellar velocities.
In order to consistently determine the impact of the Sommerfeld-enhanced $J$-factors, we must compare them against the $J$-factors represented by the gray dotted curves in Fig.~\ref{fig:rmaxvmaxgrid}.
Although it is possible to consider the impact of both non-NFW and anisotropic models, the case of anisotropic models would require extending beyond the approximation of Eq.~\eqref{eq:eddington}.

\section{Sommerfeld-Enhanced Dark Matter Annihilation}
\label{sec:sommerfeld}
\subsection{Sommerfeld Enhancement for a Yukawa Potential}

We consider two dark matter particles $X$ that interact via the exchange of a light mediator $\phi$ of mass $m_\phi$ with a coupling $g_X=\sqrt{4\pi\alpha_X}$.
For a scalar or vector mediator, the attractive force between nonrelativistic dark matter particles is described by a Yukawa potential
\begin{equation}
  V(r) = -\frac{\alpha_X}{r} e^{-m_\phi r} \ .
\end{equation}
Although dark matter annihilation can essentially be thought of as a contact interaction, the long range of the potential causes distortion of the incoming dark matter particles' wave function $\psi(\vec{r})$ (which is asymptotically a plane wave) at nonzero separation $\vec{r}$.
As a result, the annihilation cross section is enhanced by a factor $S \equiv |\psi(0)|^2$.
We write the annihilation cross section in the absence of the long-range Yukawa interaction as $(\sigma_A v_\rel)_0$, which we assume is non-vanishing in the limit $v_\rel \rightarrow 0$, where $v_\rel$ is the relative velocity of the dark matter particles.
Thus, the Sommerfeld-enhanced cross section may be written as $(\sigma_A v_\rel) = (\sigma_A v_\rel)_0 \times S$.

We briefly describe the quantum mechanics behind the Sommerfeld enhancement and refer the reader to Ref.~\cite{ArkaniHamed:2008qn} for a more detailed review.
With the central Yukawa potential, the annihilation process is determined by solving a 1D radial Schr\"{o}dinger equation for the relative motion of the dark matter particles.
We recast the physical parameters of the theory into the dimensionless quantities
\begin{equation}
  \epsilon_v \equiv \frac{v}{\alpha_X} \quad\textrm{and}\quad
  \epsilon_\phi \equiv \frac{m_\phi}{\alpha_X m_X} \ ,
\end{equation}
where $v = v_\rel/2$ is the velocity of the dark matter particles in the center-of-mass frame.
Using the dimensionless variable $x=\alpha_X m_X r$, the radial Schr\"{o}dinger equation becomes
\begin{equation}
  \chi^{\prime\prime}(x) + \left[\epsilon_v^2 + V(x) \right] \chi(x) = 0 \ ,
  \label{eq:SE}
\end{equation}
with the potential $V(x) = \exp(-\epsilon_\phi x)/x$.
We assume the annihilation is $s$-wave and neglect higher partial-wave contributions.
For the boundary conditions $\chi(x) = \exp(i\epsilon_v x)$ and $\chi'(x) = i\epsilon_v \chi(x)$ as $x\to\infty$, the Sommerfeld enhancement is $S = \left|\chi(\infty)/\chi(0)\right|^2$.

\begin{figure}[t]
  \centering
 \includegraphics[scale=0.5]{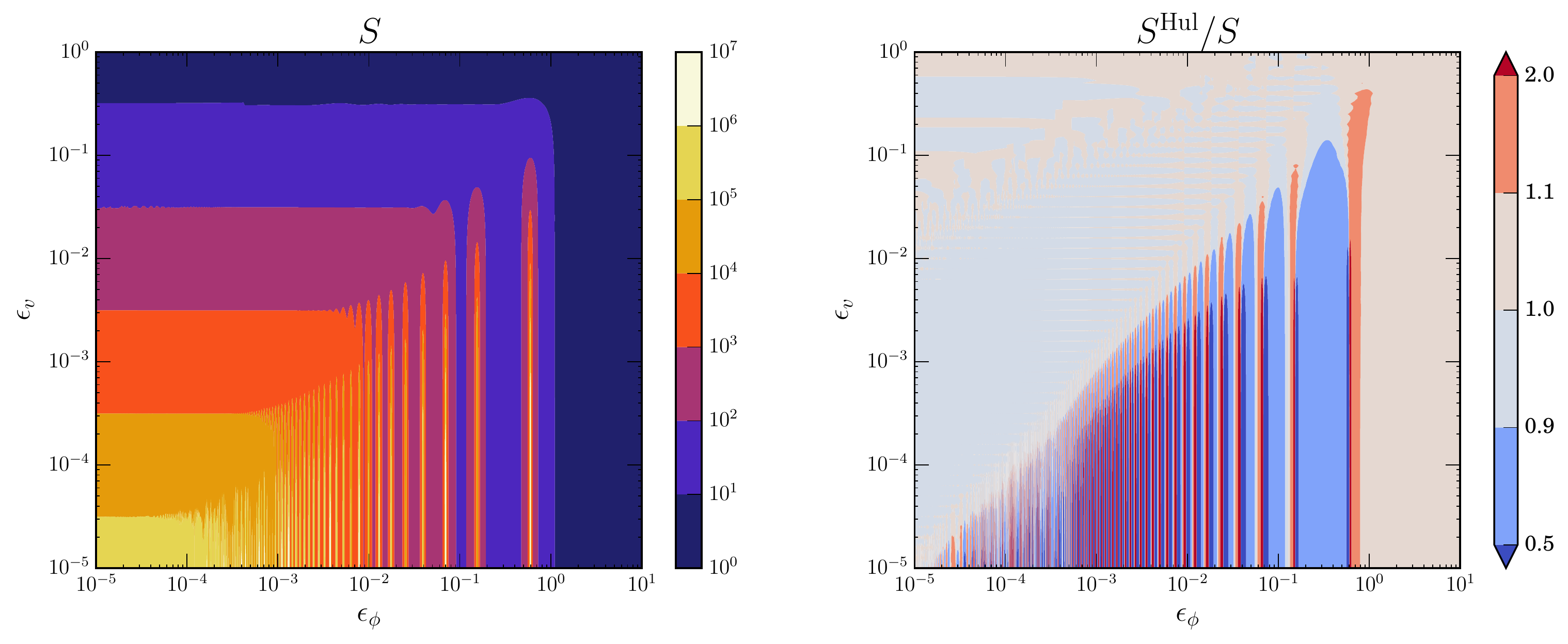}
  \caption{Contour plots of the Sommerfeld enhancement $S$ and the ratio of $S^\textrm{Hul}/S$, where $S^\textrm{Hul}$ is the analytic approximation from the Hulth\'{e}n potential.
    The analytic approximation is within 10\% of the numerical solution for $\epsilon_v > \epsilon_\phi$.
    For $\epsilon_v < \epsilon_\phi$, there are discrepancies up to a factor of 2 and beyond (up to many orders of magnitude, represented by the dark red and dark blue arrows in the color bar) that correspond to the misalignment of the resonances in the Hulth\'{e}n and Yukawa solutions.}
  \label{fig:sommerfeld}
\end{figure}

The scattering solution for the case of a Yukawa potential is not analytically solvable, although there are a variety of techniques for solving Schr\"{o}dinger equation numerically in different regimes of parameter space.
By approximating the Yukawa potential as the Hulth\'{e}n potential, the resulting Schr\"{o}dinger equation can be solved analytically, yielding the Sommerfeld enhancement~\cite{Feng:2010zp}
\begin{equation}
  S \simeq \frac{\pi}{\epsilon_v}
  \frac{\sinh \left(\frac{2 \pi \epsilon_v}{\pi^2 \epsilon_\phi / 6} \right)}
       {\cosh \left(\frac{2 \pi \epsilon_v}{\pi^2 \epsilon_\phi / 6} \right)
         - \cos \left( 2\pi \sqrt{\frac{1}{\pi^2 \epsilon_\phi / 6}
           - \frac{\epsilon_v^2}{(\pi^2 \epsilon_\phi / 6)^2} } \right)} \ .
  \label{eq:S}
\end{equation}
The analytic approximation is typically within $\sim 10\%$ of the result found from the numerical calculation, as seen in Fig.~\ref{fig:sommerfeld}.
Substantial differences for $\epsilon_v < \epsilon_\phi$ arise in narrow regions of parameter space around specific values of $\epsilon_\phi$, due to the location of the resonances of the Hulth\'{e}n potential (described below) not quite lining up to those of the Yukawa potential.
Nonetheless, the analytic approximation exhibits the same generic features as that from the full numerical solution.
As expected, in the limit where $\phi$ is heavy ($\epsilon_\phi \gg 1$), we find $S \rightarrow 1$, and there is no Sommerfeld enhancement.
In the limit $\epsilon_\phi \ll \epsilon_v$, we find $S \rightarrow \pi/ \epsilon_v = \pi \alpha_X /v$, which is the standard result for Sommerfeld enhancement in the presence of a Coulomb force.
In the limit $\epsilon_v \ll \epsilon_\phi$, we have
\begin{equation}
  S \simeq \frac{12\, \alpha_X m_X}{m_\phi} = \frac{12}{\epsilon_\phi} \ .
\end{equation}
However, in this regime, there are certain values of $\epsilon_\phi$ for which resonances occur:
\begin{equation}
  \epsilon_\phi \simeq \frac{6}{\pi^2 n^2}
  \quad \textrm{for}\ n \in \mathbb{Z}^+ \ ,
\end{equation}
where the argument of the cosine in the denominator of Eq.~\eqref{eq:S} vanishes.
Equivalently, the resonances occur for $m_\phi \simeq 6 \alpha_X m_X / (\pi^2 n^2)$, at which
\begin{equation}
  S \simeq \frac{\alpha_X^2}{v^2 n^2} = \frac{1}{\epsilon_v^2 n^2} \ .
\end{equation}

In the limit of $v\to 0$, the resonant enhancements become large and unphysical, because we have thus far neglected the effects of zero-energy bound-state formation and decay~\cite{Hisano:2004ds,Feng:2010zp}.
By inserting a $\delta$-function, which is sufficient for $s$-wave processes, into the Schr\"{o}dinger equation to represent the short-range interaction, these resonances are regularized and yield cross sections that obey partial-wave unitarity bounds~\cite{Blum:2016nrz}.
For a perturbative, short-range annihilation cross section $(\sigma_A v_\rel)_0 \ll 4\pi/m_X^2 v_0$, the standard Sommerfeld enhancement $S$ is modified as follows:
\begin{equation}
  \tilde{S}(v) = \frac{S(v)}{\left|1-i\epsilon_v\alpha_X
    \frac{m_X^2}{8\pi}(\sigma_A v_\rel)_0 \left[T(v)+iS(v)\right]\right|^2} \ ,
  \label{eq:Smod}
\end{equation}
where $T$ is another quantity that encodes the effect of the long-range Yukawa force on the wave function and depends on the renormalization of the $\delta$-function.
We have neglected the real part of the inverse-scattering length, which corresponds to setting the short-range scattering cross section to be $\sigma_{\textrm{sc},0} = {(\sigma_A v_\rel)_0}^2 (m_X/2)^2 / (4\pi)$.
Such an identification may arise in a nonrelativistic theory in which the optical theorem relates the annihilation cross section to the imaginary part of the forward-scattering amplitude~\cite{Hisano:2002fk}.
For $\alpha_X \ll 1$, the denominator in Eq.~\eqref{eq:Smod} approaches 1, except for the region very close to the resonance.
We have verified that for small $\alpha_X$, the correction due to bound states is essentially limited to the peak of the resonance but has little effect otherwise.
For our main analysis, we set $\alpha_X=10^{-2}$ and simply use $S$, allowing us to avoid the issue of model-dependence in choosing the form of $(\sigma_A v_\rel)_0$.

It is worth noting that, around the epoch of recombination, it is expected that the typical dark matter particle velocity would be much smaller than it is in the current epoch.
One should thus worry that models with Sommerfeld-enhanced annihilation that could potentially be observed with future experiments would already be ruled out by constraints from the Plank experiment on dark matter annihilation in the early Universe~\cite{Ade:2015xua}, at which time the Sommerfeld-enhancement could be much larger (see, for example,~\cite{Finkbeiner:2010sm}).
Constraints on dark matter annihilation in the early Universe will not be enhanced, relative to constraints arising from observations of dSphs, provided the Sommerfeld enhancement saturates for velocities not far below the typical dark matter velocity in a dSph, which is ${\cal O}(1)~{\rm km} /{\rm s}$.
This is equivalent to the constraint $\epsilon_\phi \gtrsim 10^{-6} \alpha_X^{-1}$.

\subsection{Relating the Sommerfeld Enhancement to the Photon Flux}

If the dark matter particle is its own anti-particle, the differential photon flux produced by dark matter annihilation is
\begin{equation}
  \frac{d\Phi}{dE_\gamma} =
  \frac{1}{4\pi} \frac{dN}{dE_\gamma} \int_{\Delta\Omega} d\Omega \int d\ell
  \int d^3 v_1 \frac{f(r(\ell, \Omega), \vec{v}_1)}{m_X}
  \int d^3 v_2 \frac{f(r(\ell, \Omega), \vec{v}_2)}{m_X}
  \, \frac{(\sigma_A |\vec{v}_1 - \vec{v}_2|) }{2} \ ,
\end{equation}
where $\ell$ is the distance along the line of sight and $dN/dE_\gamma$ is the photon spectrum produced by a single annihilation process.
The angular integration over $\Delta\Omega$ covers a region in the sky encompassing a particular dSph.
If the dark matter particle and anti-particle are distinct and equally abundant, this flux would be suppressed by an additional factor of $1/2$.

Expressing the Sommerfeld-enhanced annihilation cross section as $(\sigma_A v_\rel) = (\sigma_A v_\rel)_0 S(v_\rel/2)$, where $v_\rel = |\vec{v}_1 - \vec{v}_2|$, we have
\begin{equation}
  \frac {d\Phi}{dE_\gamma} = J_S(\Delta\Omega)
  \frac{(\sigma_A v_\rel)_0}{8\pi m_X^2} \frac{dN}{dE_\gamma} \ ,
\end{equation}
where
\begin{equation}
  J_S (\Delta\Omega) \equiv \int_{\Delta\Omega} d\Omega \int d\ell
  \int d^3v_1 f(r(\ell, \Omega), \vec{v}_1)
  \int d^3v_2 f(r(\ell, \Omega), \vec{v}_2) \,
  S(|\vec{v}_1-\vec{v}_2|/2)
\end{equation}
is the Sommerfeld-enhanced $J$-factor, which encapsulates all of the dependence of the photon flux on the dark matter distribution of the target.
Note that since $S$ is a function of the velocity $|\vec{v}_1-\vec{v}_2|/2$, it depends on the angle between $\vec{v}_1$ and $\vec{v}_2$, as well as the magnitudes $v_1$ and $v_2$.
In the limit $S \rightarrow 1$ (\textit{i.e.}, no Sommerfeld enhancement) we recover the ordinary result
\begin{equation}
  J_S (\Delta\Omega) \rightarrow J(\Delta\Omega) =
  \int_{\Delta\Omega} d\Omega \int d\ell \left[\rho(r(\ell, \Omega))\right]^2 \ .
\end{equation}
Our main goal is to determine $J_S (\Delta \Omega)$ for a variety of dwarf spheroidal galaxies.
In general, $J_S$ depends on two parameters of the particle physics model: $\epsilon_\phi$ and $\alpha_X$.

\section{Results}
\label{sec:results}
\subsection{Sommerfeld-enhanced $J$-factors}

\begin{figure}[t]
  \centering
  \includegraphics[scale=0.6]{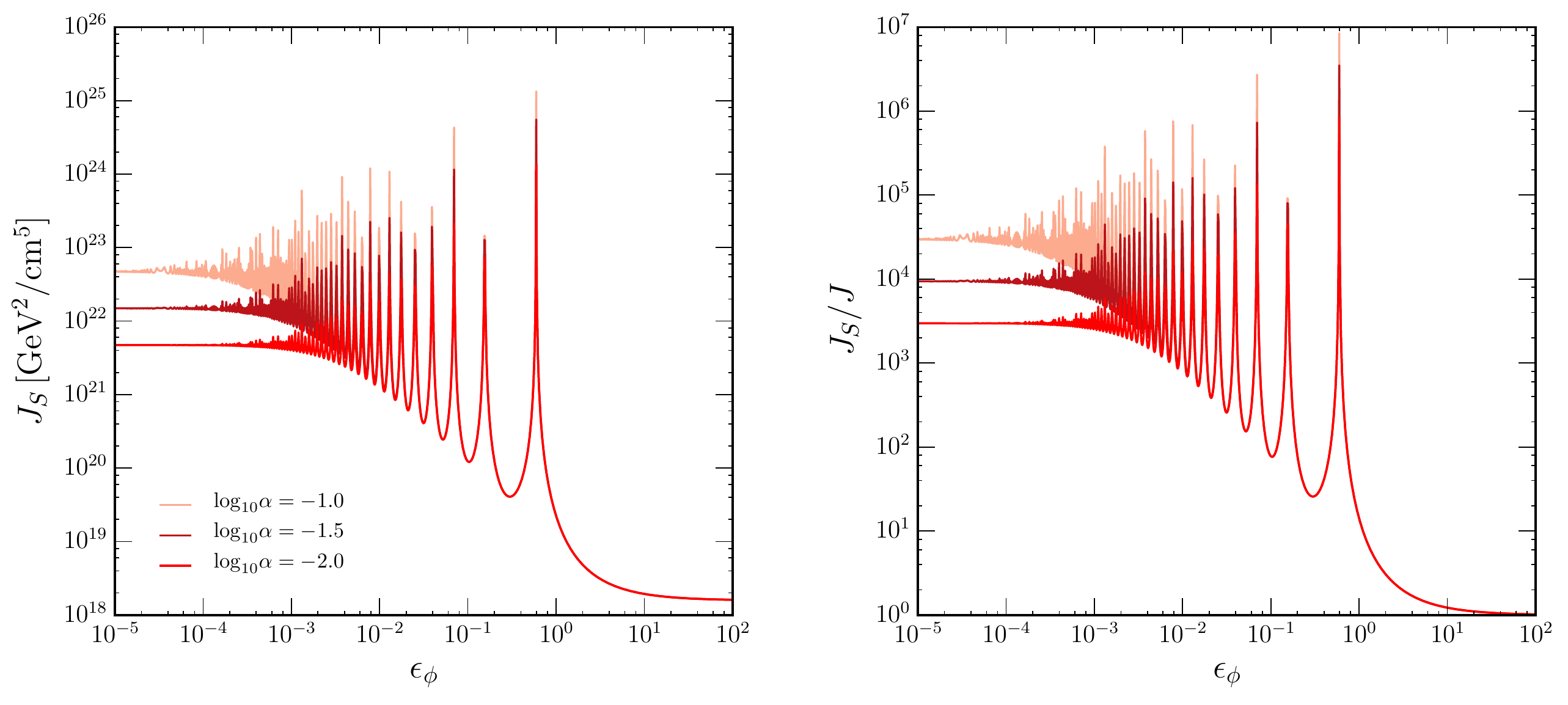}
  \caption{Sommerfeld-enhanced $J$-factors for Reticulum II, using the central point in Fig.~\ref{fig:rmaxvmaxgrid}.
    We vary $\alpha_X$ over an order of magnitude and choose $\Delta\Omega \approx 2.4 \times 10^{-4}$, corresponding to a cone with a half-angle of $0.5^\circ$.
    The left panel shows the values of $J_S$, while the right panel shows the ratio of $J_S$ to the ordinary $J$-factor calculation with no enhancement.}
  \label{fig:jfactor_alpha}
\end{figure}

We calculate the Sommerfeld-enhanced $J$-factor for five of the most promising dSphs for indirect detection: Segue 1, Reticulum II, Coma Berenices, Draco, and Ursa Minor.
In Fig.~\ref{fig:jfactor_alpha}, we plot $J_S$ as a function of $\epsilon_\phi$ for Reticulum II for different values of $\alpha_X$, assuming $\Delta \Omega = 2.4 \times 10^{-4}$ (\textit{i.e.}, a cone half-angle of $0.5^\circ$) and nominal values of the NFW parameters given in Table~\ref{tb:orbit}.
As expected, away from the resonances, we find that $J_S$ scales as $\alpha_X$.
But near resonances, $S$ scales as $\alpha_X^2 \epsilon_\phi / v^2$; the magnitude of the resonances are thus suppressed for small $\epsilon_\phi$ and essentially disappear for small $\epsilon_\phi$ and $\alpha_X$.
In accordance with Sec.~\ref{sec:sommerfeld}, we henceforth focus on the benchmark case $\alpha_X = 10^{-2}$.

\begin{figure}[t]
  \centering
  \includegraphics[scale=0.6]{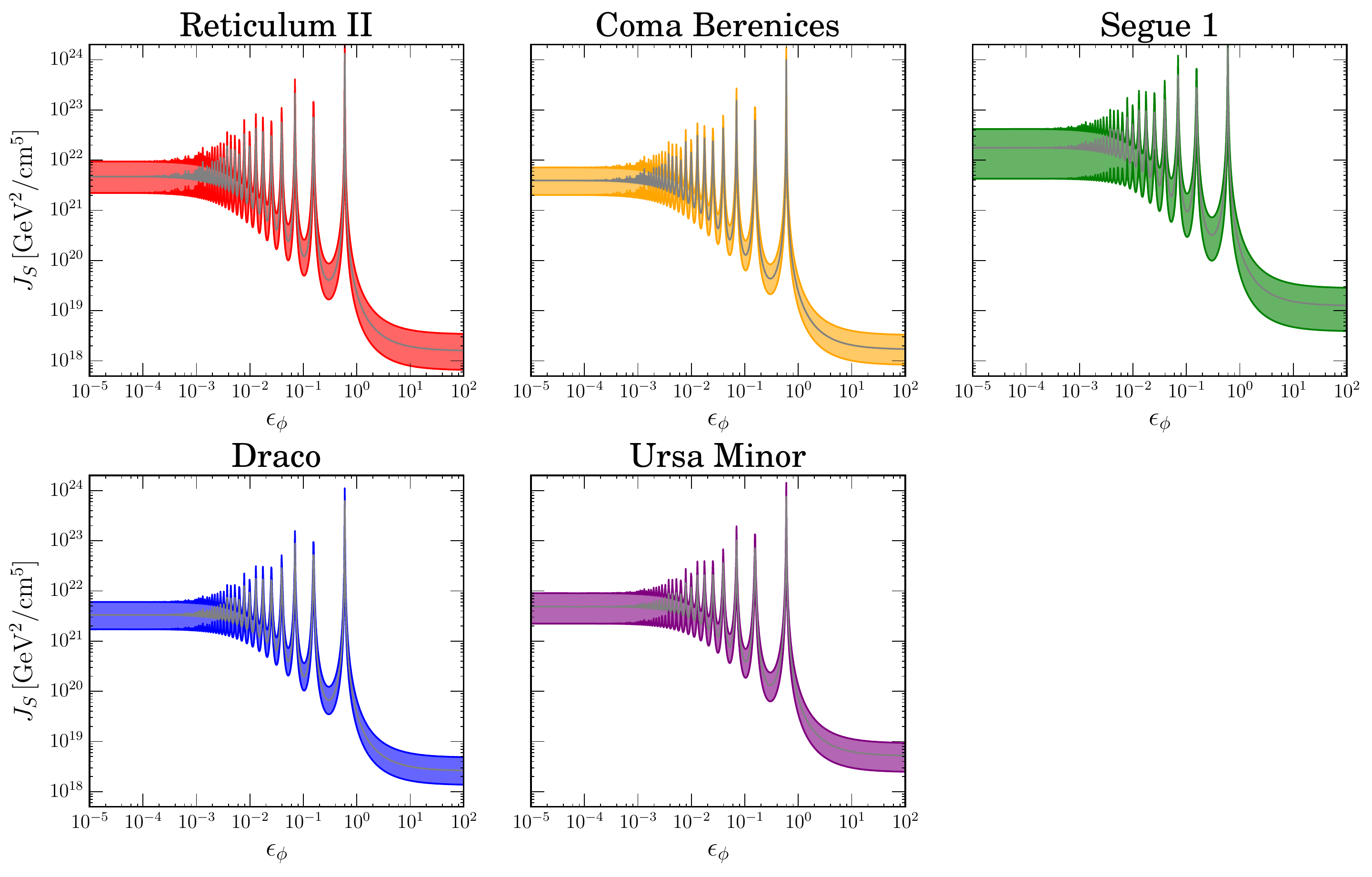}
  \caption{Sommerfeld-enhanced $J$-factors with $\alpha=10^{-2}$ and $\Delta\Omega \approx 2.4 \times 10^{-4}$, corresponding to a cone half-angle of $0.5^\circ$.
    The width of the bands represents the systematic uncertainties we estimate by using the points in Fig.~\ref{fig:rmaxvmaxgrid}.
    The NFW parameters for the gray lines represent the central point.}
  \label{fig:jfactor_all}
\end{figure}

In Fig.~\ref{fig:jfactor_all}, we plot $J_S$ as a function of $\epsilon_\phi$ for all five dSphs, assuming $\alpha_X = 10^{-2}$ and $\Delta \Omega = 2.4 \times 10^{-4}$.
The width of the bands show the uncertainty in choosing the NFW parameters, as defined by the five points in Fig.~\ref{fig:rmaxvmaxgrid}.
The gray line shows the central point for each dSph, which represents the NFW parameters that match both the average stellar velocity dispersion and the median ($\Vmax,\rmax$) relation from simulations.
As expected, $J_S \rightarrow J$ in the limit $\epsilon_\phi \gg 1$.
However, $J_S / J \sim 10^3$ for small $\epsilon_\phi$, while near resonances $J$ and $J_S$ differ by many orders of magnitude; this corresponds to the factor by which sensitivity to $(\sigma_A v_\rel)_0$ is improved by Sommerfeld enhancement.

\begin{figure}[t]
  \centering
  \includegraphics[scale=0.6]{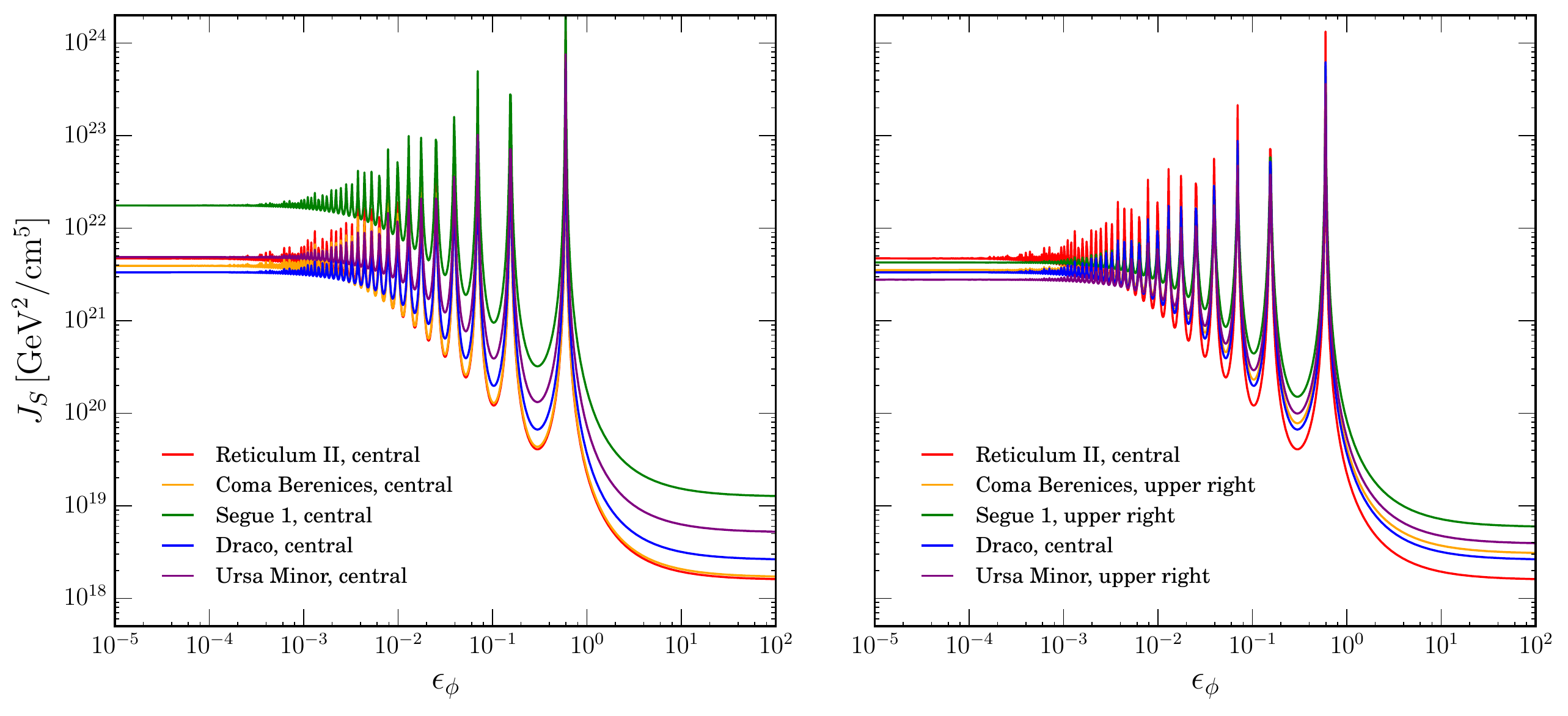}
  \caption{Comparison of Sommerfeld-enhanced $J$-factors for the five dSphs.
    In the left panel, we use the central points in Fig.~\ref{fig:rmaxvmaxgrid}, which are also represented by the gray lines in Fig.~\ref{fig:jfactor_all}.
    In the right panel, we use the central points in Fig.~\ref{fig:rmaxvmaxgrid} for Reticulum II and Draco and the upper-right points in Fig.~\ref{fig:rmaxvmaxgrid} [i.e., points with the largest ($\Vmax$, $\rmax$)] for the remaining dSphs.
    This specific combination of NFW parameters for the right panel is a concrete example of Reticulum II having the smallest relative $J_S$-factor at large $\epsilon_\phi$, but the largest relative $J_S$-factor at small $\epsilon_\phi$.}
  \label{fig:jfactor_cmp}
\end{figure}

It is interesting to note that the relative order of $J_S$ among the dSphs may change as a function of $\epsilon_\phi$.
In particular, in the limit of no Sommerfeld enhancement ($\epsilon_\phi \gg 1$), Reticulum II tends to have a smaller $J$-factor than the other dSphs; however, in the limit of $\epsilon_\phi \ll 1$, the $J_S$-factor for Reticulum II seems relatively higher in comparison.
In the left panel of Fig.~\ref{fig:jfactor_cmp}, we show the $J_S$-factors from the gray lines in Fig.~\ref{fig:jfactor_all}, plotted together for ease of comparison.
To emphasize how the astrophysical uncertainty affects the relative ordering, we choose particular points from those listed in Fig.~\ref{fig:rmaxvmaxgrid} to show in the right panel of Fig.~\ref{fig:jfactor_cmp}.
These points represent a scenario in which Reticulum II has the smallest ordinary $J$-factor at large $\epsilon_\phi$, but has the largest $J_S$-factor at small $\epsilon_\phi$.
Moreover, Reticulum II maintains its status of having the lowest $J_S$ factor in the valleys between resonance peaks, but settles into a higher $J_S$ factor once $\epsilon_\phi$ is small enough and away from the resonant regime.

Although we have chosen a particular value of $\alpha_X$, the relative order of $J_S$ among the dSphs (for a given set of NFW parameters) is unaffected for a different value of $\alpha_X$ at small and large $\epsilon_\phi$.
The $J_S$ dependence on $\alpha_X$ is fairly straightforward, as described at the beginning of this section: for $\epsilon_{\phi} \gtrsim 1$, $J_S$ is independent of $\alpha_X$, while for $\epsilon_{\phi} \ll 1$ (but away from resonances), one instead finds $J_S \propto \alpha_X$.
Thus, changing the value of $\alpha_X$ scales $J_S$ for all dSphs by the same amount, resulting in no change in relative ordering among the dSphs outside of the resonant regime.

\subsection{An Analytic Approximation to the Determination of $J$ and $J_S$}

Interestingly, it is possible to use analytic results to generate simple expressions which determine if the relative order of $J_S$ at small $\epsilon_\phi$ for two dSphs is different from the relative order of $J$.
For this purpose, we focus on comparing two limits: the non-enhanced limit ($\epsilon_\phi \gg 1$) and the limit of a Coulomb-like potential ($\epsilon_\phi \ll 1$).

We assume that the density profile may be expressed in the form
\begin{equation}
  \rho (r) = \rho_s  \times \tilde \rho ( r / r_s) \ ,
  \label{eq:rho_form}
\end{equation}
where $\rho_s $ is an overall density scale, and $\tilde \rho$ is a dimensionless quantity which may be expressed as a function of $\tilde r \equiv r / r_s$ only.
Thus, $\tilde \rho (\tilde r)$ is a dimensionless function which is independent of the parameters that characterize the dark matter distribution.
For the particular case of an NFW profile, we have $\tilde \rho (\tilde r) = \tilde r^{-1} (1+\tilde r)^{-2}$.

If the integration over a solid angle $\Delta \Omega$ is large enough to essentially encompass the entire region of the dSph in which there is significant dark matter annihilation, the $J$-factor can be expressed in terms of an integral over the radial distance from the center of the dwarf, instead of an integral over the line of sight:
\begin{equation}
  J^\textrm{total} = \frac{1}{D^2} \int dV \, [\rho (r)]^2
  = \frac{4\pi}{D^2} \int dr \, r^2 [\rho (r)]^2 \ ,
\end{equation}
where we have assumed that $\rho (r)$ is negligible unless $r \ll D$.
We then find
\begin{subequations}
  \begin{align}
    J^\textrm{total} &= \frac {4\pi \rho_s^2 r_s^3}{D^2} C_J \\
    C_J &\equiv \int d\tilde r \, \tilde r^2 [\tilde \rho (\tilde r)]^2 \ ,
  \end{align}
\end{subequations}
where $C_J$ is a dimensionless quantity that depends on the functional form of the dark matter distribution, but is independent of the parameters ($\rho_s$, $r_s$).
For an NFW profile, $C_J = 1/3$.

To determine the Sommerfeld-enhanced $J$-factor, we express the dark matter velocity distribution in a scale-invariant form by utilizing the fact that the gravitational potential can be written as
\begin{subequations}
  \begin{align}
    \Psi (r) &= G \rho_s r_s^2 \times \tilde \Psi (\tilde r) \\
    \tilde \Psi (\tilde r) &\equiv \int_\infty^{\tilde r} dx \, \frac{1}{x^2}
    \int_0^x  dy \, (4\pi y^2) \tilde \rho (y) \ ,
  \end{align}
\end{subequations}
where $\tilde{\Psi}$ is a dimensionless function that is independent of the parameters of the dark matter distribution.
We define a scale-invariant velocity $\tilde v \equiv (G \rho_s r_s^2)^{-1/2} v$ and a scale-invariant energy per unit mass $\tilde \epsilon = \tilde v^2 /2 + \tilde \Psi$.
In terms of these quantities, the dark matter distribution function may be rewritten as
\begin{subequations}
  \begin{align}
    f_\DM (\epsilon)
    &= \rho_s (G \rho_s r_s^2)^{-3/2} \tilde f_\DM (\tilde \epsilon) \\
    \tilde f_\DM (\epsilon)
    &\equiv \frac{1}{\sqrt{8}\pi^2} \int_{\tilde \epsilon}^0
    \frac{d^2 \tilde \rho}{d \tilde \Psi^2}
    \frac{d \tilde \Psi}{\sqrt{\tilde \epsilon - \tilde \Psi}} \ ,
  \end{align}
\end{subequations}
where $\tilde{f}_\DM$ is also a dimensionless quantity that is independent of ($\rho_s$, $r_s$), but is implicitly a function of $\tilde r$ and $\tilde v$.

In the limit $\epsilon_\phi \ll 1$, we approximate the Sommerfeld enhancement factor by $S \sim \pi \alpha_X / v$; that is, we assume the contribution to dark matter annihilation arising from the region of phase space with $\epsilon_v \lesssim \epsilon_\phi$ is negligible.
By integrating over an angle which encompasses the entire dwarf, we can write $J_S$ as
\begin{subequations}
  \begin{align}
    {J_S}^\textrm{total} &\sim
    \left(\frac{4\pi \rho_s^2 r_s^3}{D^2}\right) (G\rho_s r_s^2)^{-1/2} C_{J_S} \\
    C_{J_S} &\equiv \int d\tilde r \, \tilde r^2
    \int d^3 \tilde v_1 \tilde f (\tilde r, \tilde v_1)
    \int d^3 \tilde v_2 \tilde f (\tilde r, \tilde v_2)
    \left(\frac{2\pi \alpha_X}{|\overrightarrow{\tilde v}_1
      -\overrightarrow{\tilde v}_2|} \right) \ ,
  \end{align}
\end{subequations}
where $C_{J_S}$ is a dimensionless quantity that depends on the functional form of the dark matter distribution, but not on the parameters ($\rho_s$, $r_s$).

In summary, we find
\begin{align}
  J^\textrm{total} &\propto \rho_s^2 r_s^3 / D^2 \ , \nonumber\\
  {J_S}^\textrm{total} &\propto \rho_s^{3/2} r_s^2 / D^2 \ .
  \label{eq:j_js_join}
\end{align}
Thus, for fixed $J^\textrm{total}$, one increases $J_S^\textrm{total}$ by increasing $\rho_s$ and correspondingly decreasing $r_s$.
In other words, if two dSphs have distributions with the same functional form and have the same total $J$-factor, then the dwarf with the smaller scale size will have the larger Sommerfeld-enhanced $J$-factor (in the limit $\epsilon_\phi \ll 1$).

It is important to note that these results do not depend on the assumption of an NFW profile; they apply for any choice of density profile, provided the radial dependence can be expressed entirely in terms of the dimensionless variable $\tilde r = r/r_s$.
This result does depend on the assumption that the solid angle encompasses the entire dSph.
There are several NFW parameter choices we have studied for which this assumption is not true, implying that the specific results we have found for $\Delta \Omega \approx 2.4 \times 10^{-4}$ need not obey these scaling relations precisely.
Nevertheless, they provide useful guidance for the general criteria governing the situations in which the ordering of Sommerfeld-enhanced $J$-factors differs from that of non-enhanced $J$-factors.

\begin{figure}[t]
  \centering
  \includegraphics[scale=0.4]{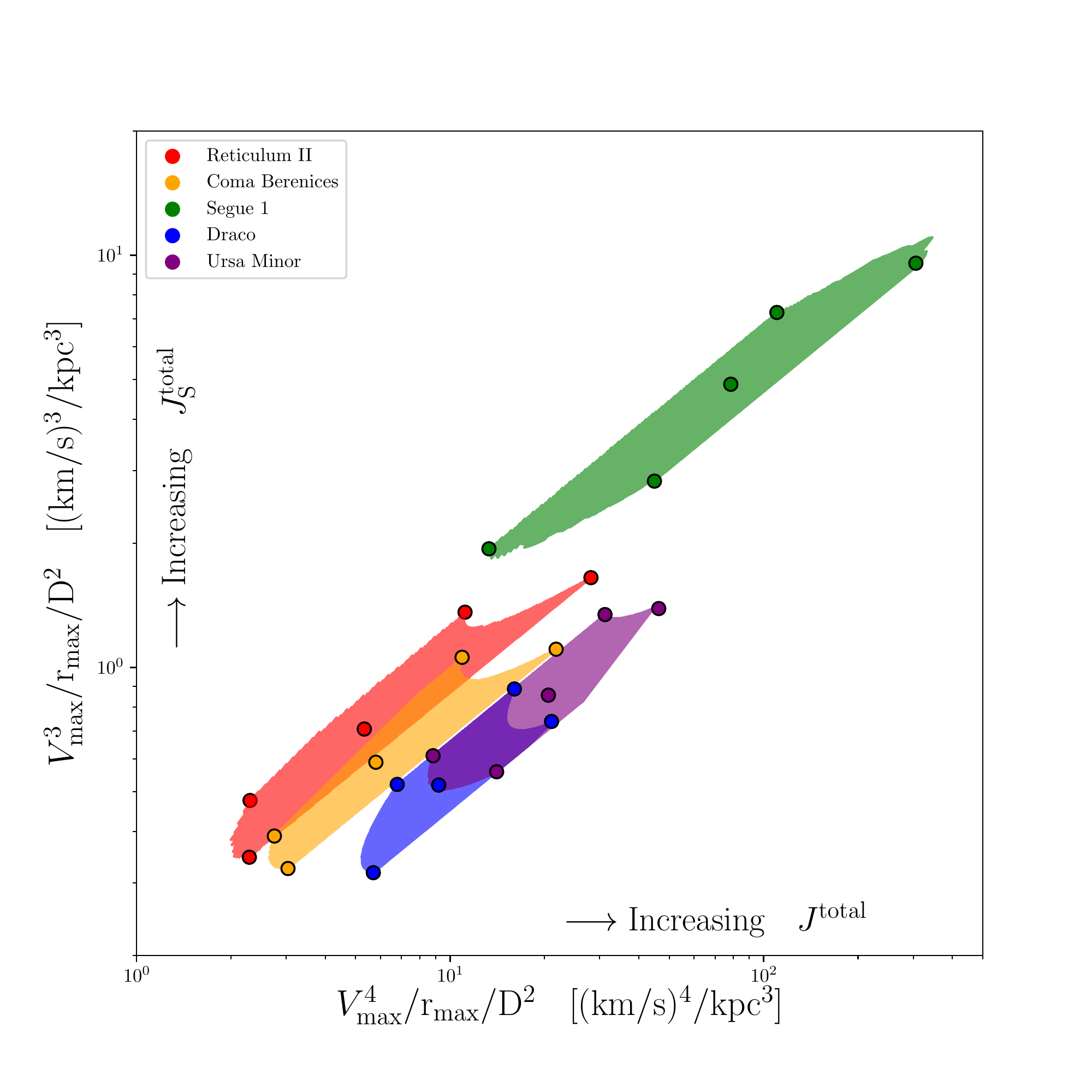}
  \caption{Regions in the $J^\textrm{total}$--${J_S}^\textrm{total}$ parameter space, as defined in Eq.~\ref{eq:j_js_join}, for each of the dSphs.
    The boundaries of each region, and the respective central points, are defined in Fig.~\ref{fig:rmaxvmaxgrid}.}
  \label{fig:j_js_join}
\end{figure}

With this in mind, in Fig.~\ref{fig:j_js_join} we plot the NFW profile parameter space that we have considered for each of the five dSphs.
The shaded region for each dSph encompasses the five benchmark NFW parameters shown in Fig.~\ref{fig:rmaxvmaxgrid}.
The parameter space shown is $(\Vmax^4/\rmax/D^2, \Vmax^3/\rmax/D^2)$, which is equivalent to the parameter space of interest, $(\rho_s^2 r_s^3 / D^2 , \rho_s^{3/2} r_s^2 / D^2)$ from Eq.~\eqref{eq:j_js_join}.
This relation between parameters is clear from dimensional analysis: we have $\rmax \propto r_s$ from the assumed form of $\rho(r)$ in Eq.~\eqref{eq:rho_form}; and since $\Vmax^2 = GM(r=\rmax)/\rmax$, where the mass function $M(r)$ merely involves integrating over $\rho$, we find that $\Vmax^2 \propto G \rho_s r_s^2$.
Parameter space points yield larger values of $J^\textrm{total}$ as one moves to the right in Fig.~\ref{fig:j_js_join}, and larger values of $J_S^\textrm{total}$ (assuming $\epsilon_\phi \ll 1$) as one moves up.
Thus, if there exists a parameter point for one dSph which lies above and to the left of a parameter point of another dSph, then for those choices of parameters, the ordering of $J^\textrm{total}$ will differ from that of $J_S^\textrm{total}$; the point to the upper left will have a smaller $J^\textrm{total}$, but a larger $J_S^\textrm{total}$.

As illustrated by Figure~\ref{fig:j_js_join}, there are several pairs of dSphs for which the relative ordering of $J$ can be different from the relative ordering of $J_S$ in the limit $\epsilon_\phi \ll 1$.
In particular, there are choices of NFW parameters for which Reticulum II has a smaller $J^\textrm{total}$ than either Coma Berenices, Draco, or Ursa Minor; but has a larger $J_S^\textrm{total}$ than all of the others.
But there is no choice of parameters for which Reticulum II has a larger $J_S^\textrm{total}$ than Segue 1.
However, for Segue 1, the consistent region of parameter space includes points with a relatively large ($\Vmax$, $\rmax$); such points not only yield a relatively small $J_S^\textrm{total} / J^\textrm{total}$, but also a large angular size (especially since Segue 1 is the closest of these five dSphs).
Thus, although Segue 1 may always have a larger $J_S^\textrm{total}$ than Reticulum II, there are points in parameter space for which Reticulum II will have the larger $J_S$ (in the $\epsilon_\phi \ll 1$ limit) when integrated over a cone of half-angle $0.5^\circ$, as shown in Figure~\ref{fig:jfactor_cmp}.

\section{Conclusion}
\label{sec:conclusion}

In this paper we have self-consistently calculated dSph $J$-factors, using a model for the dark matter phase space distribution and gravitational potentials that are constrained from stellar kinematics.
Within the context of our spherically-symmetric and isotropic model for the dark matter phase space distribution, we quantify the astrophysical uncertainty in the $J_S$-factors and show that the relative ordering of the most promising $J_S$-factors can be interchanged relative to the standard velocity-independent $J$-factors.
This result may have important implications for the interpretation of possible gamma-ray excesses from a dSph~\cite{Geringer-Sameth:2015lua,Li:2015kag}.

The model that we discuss can be seen as a first step in the self-consistent calculation of astrophysical $J$-factors for velocity-dependent annihilation cross sections.
A new step in this analysis could involve the determination of $J$-factors for other types of models with a velocity-dependent annihilation cross section~\cite{Zhao:2016xie}.
Although certain forms of the velocity dependence, such as $\sigma_A v \propto v^2$, reduce the cross section below the sensitivity of current experiments, these models may become accessible as the sensitivity improves.
This will be particularly true as new dSphs continue to be discovered by the Dark Energy Survey~\cite{Drlica-Wagner:2015ufc}, and even further into the future by the Large Synoptic Survey Telescope.
Accounting for velocity-dependent cross sections will also be a future requirement of numerical methods that scan the theoretical parameter space given large spectroscopic data sets~\cite{Bonnivard:2015pia,Chiappo:2016xfs}.

From a theoretical perspective, our analysis may be improved by extending beyond NFW profiles and isotropic stellar velocity dispersions.
The dark matter potentials of dSphs are at present consistent with both cores~\cite{Walker:2011zu} and cusped profiles~\cite{Breddels:2013qqh,Richardson:2013lja,Strigari:2014yea}, and the shape of the dark matter phase space distribution is different in both cases, even for isotropic models.
For anisotropic models, the dark matter distribution function depends on additional integrals of motion beyond just the energy.
Guidance may come from cosmological simulations, which are able to determine the phase space distribution of the dark matter and stars separately~\cite{Campbell:2016vkb}.

\bigskip
{\bf Acknowledgements}

KB and JK are supported in part by NSF CAREER grant PHY-1250573. LES acknowledges support from NSF grant PHY-1522717.
We would also like to acknowledge the Center for Theoretical Underground Physics and Related Areas (CETUP$^\ast$) for hospitality and partial support during the 2016 Summer Program.


\end{document}